# SAPO-RL: <u>S</u>equential <u>A</u>ctuator <u>P</u>lacement <u>O</u>ptimization for Fuselage Assembly via <u>R</u>einforcement <u>L</u>earning

Peng Ye [a], Juan Du [abc]∗

[a] IDADM Lab, Smart Manufacturing Thrust, The Hong Kong University of Science and Technology (Guangzhou), Guangzhou 511455, China
[b] Data Science and Analytics Thrust, The Hong Kong University of Science and Technology (Guangzhou), Guangzhou 511455, China
[c] The Hong Kong University of Science and Technology, Hong Kong SAR 999077, China
*Corresponding Author, Juan Du, juandu@ust.hk

## Abstract

Precise assembly of composite fuselages is critical for aircraft assembly to meet the ultra-high precision requirements. Due to dimensional variations, there is a gap when two fuselage assemble. In practice, actuators are required to adjust fuselage dimensions by applying forces to specific points on fuselage edge through pulling or pushing force actions. The positioning and force settings of these actuators significantly influence the efficiency of the shape adjustments. The current literature usually predetermines the fixed number of actuators, which is not optimal in terms of overall quality and corresponding actuator costs. However, optimal placement of actuators in terms of both locations and number is challenging due to compliant structures, complex material properties, and dimensional variabilities of incoming fuselages. To address these challenges, this paper introduces a reinforcement learning (RL) framework that enables sequential decision-making for actuator placement selection and optimal force computation. Specifically, our methodology employs the Dueling Double Deep Q-Learning (D3QN) algorithm to refine the decision-making capabilities of sequential actuator placements. The environment is meticulously crafted to enable sequential and incremental selection of an actuator based on system states. We formulate the actuator selection problem as a submodular function optimization problem, where the sub-modularity properties can be adopted to efficiently achieve near-optimal solutions. The proposed methodology has been comprehensively evaluated through numerical studies and comparison studies, demonstrating its effectiveness and outstanding performance in enhancing assembly precision with limited actuator numbers.

# 1. Introduction

In recent years, advanced composite materials have gained widespread application in large space structures, attributed to their exceptional characteristics such as a high strength-to-weight ratio. During the manufacturing process, inherent dimensional discrepancies may arise in fuselage assembly due to variations in manufacturing batches or suppliers. When two fuselages are joined together, a gap emerges, as depicted in Figure 1. This gap can significantly impact the assembly quality and efficiency of the fuselage assembly process. Consequently, shape adjustments at the interface between the two fuselages are essential prior to the composite fuselage assembly. In practical scenarios, actuators are employed to facilitate these shape adjustments, as illustrated in Figure 2, with actuators positioned at points from A to J. These actuators are capable of adding forces to either pull or push the fuselage at corresponding placement points. A more in-depth understanding of the utilization of actuators in the shape adjustment of composite fuselages can be referred to Wen et al. (2018) and Yue et al. (2018).

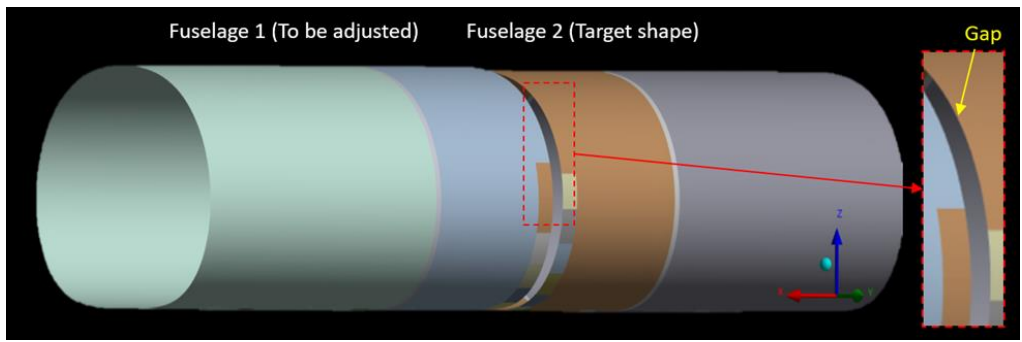

**Figure 1.** Composite fuselage assembly

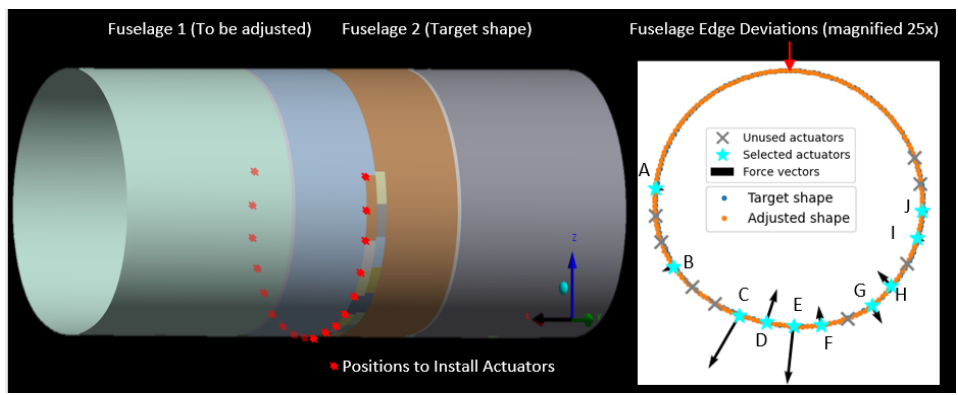

**Figure 2.** Fuselage shape adjustment by using actuators

Current practices for actuator placement are not optimal. Actuators are usually placed at equal distances between adjacent points, which may result in larger forces being applied than necessary at some locations. This not only affects the efficiency of shape adjustments but also increases the cost and complexity of the assembly process. To address these challenges, recent research has developed advanced methodologies such as surrogate model-based control strategies (Yue et al., 2018) and sparse learning-based optimal actuator placements (Du et al., 2022; Du et al., 2019). In recent years, Reinforcement Learning (RL) has received significant attention for its ability to achieve super - human performance in games. Meanwhile, some recent studies have attempted to apply RL to solve optimization problems in engineering (Hu et al., 2017; Li et al., 2024). In the past few years, there have also been relevant research efforts in optimal layout of ship hull panel fixtures (Wang et al., 2024) and fuselage actuator placement (Lutz et al., 2024). These studies all demonstrate the unique strengths of Reinforcement Learning (RL) when it comes to optimizing engineering problems.

In the current state of actuator layout optimization, a notable limitation exists: the majority of existing methodologies assume a fixed and predefined number of actuators for optimization. This constraint significantly limits the potential for holistic optimization, as the quantity of actuators directly impacts system performance, cost efficiency, and adaptability. Unlike scenarios where actuator placement can be refined through iterative adjustments, the absence of systematic strategies to optimize the number of actuators leaves a critical gap in the optimal decision-making process. Current approaches often rely on heuristic assumptions or incremental adjustments, which fail to address the interplay between actuator quantity and overall system effectiveness. As a result, practitioners are restricted to work within pre-determined parameters, potentially overlooking opportunities for more innovative and resource-efficient configurations. Addressing this limitation would require integrating actuator quantity optimization into the broader framework of layout design, enabling more flexible and adaptive solutions tailored to specific application requirements.

To address the limitations in existing actuator placement optimization works, we aim to develop a flexible reinforcement learning framework capable of sequentially selecting actuator positions and solving actuator forces in a unified manner. While this framework holds significant potential for improving optimization efficiency and adaptability, achieving this goal is challenging due to the following aspects: (i) Reinforcement learning agents require extensive experimental validation to converge to optimal policies. However, the computational load associated with repeatedly invoking finite element analysis (FEA) solvers for deformation results is prohibitively high. This bottleneck arises from the high computational cost of FEA. (ii) The problem's inherent complexity complicates the design of a suitable Markov decision process (MDP). The actuator placement optimization problem involves a large state space, dynamic interactions between actuator positions and forces, and non-linear system responses. These factors make it difficult to define a compact and effective state representation, as well as a reward function that balances exploration and exploitation while ensuring convergence to globally optimal solutions. (iii) The sequential nature of actuator selection introduces additional complexity. Each decision step impacts subsequent states, creating dependencies that require careful handling to avoid suboptimal policies. This challenge is further exacerbated by the need to balance exploration of new actuator configurations with exploitation of previously learned policies, especially in high-dimensional action spaces.

To address the challenges mentioned above, we propose a novel approach for actuator placement optimization using a flexible reinforcement learning framework. To tackle the first challenge of high computation load due to extensive experimental validation required by reinforcement learning agents and the time-consuming FEA solver invocations, we utilize a surrogate modeling technique. This technique approximates the FEA results with a reduced-order model, significantly cutting down the computational time while maintaining a reasonable level of accuracy for the deformation calculations. This allows for more efficient agent training and decision-making within practical timeframes. The key contributions are summarized as follows:

- For the second challenge of designing a suitable Markov decision process (MDP) for the complex actuator placement optimization problem, we leverage the submodular function modeling. By transforming the problem into a submodular function form, we can take advantage of its desirable properties, such as the diminishing returns property. This transformation not only simplifies the problem-solving process but also provides a solid theoretical foundation for optimization. We then convert the submodular function problem into a linear programming problem, which can be efficiently solved using well-established algorithms. This approach allows us to handle the large state space and dynamic interactions in a more manageable way.

- To address the issue of the diminishing returns in rewards $r_t$ and the decreasing norms of state representations $\boldsymbol{s}_t$ as $t$ increases, which could lead to difficulties in gradient calculation, we implement a normalization strategy. Additionally, we design a reward shaping mechanism that considers the relative improvement in the objective function instead of the absolute value. This helps maintain a stable and informative reward throughout the learning process, facilitating more effective policy learning and adaptation.

This framework not only enhances computational efficiency but is flexible enough to accommodate a variety of optimization objectives. It can not only select a certain number of actuators to provide near-optimal shape adjustments but also minimize the number of actuators under certain manufacturing requirements. The proposed methodology has been comprehensively evaluated through numerical case studies and comparison studies, demonstrating its effectiveness in enhancing assembly precision.

The remainder of this article is organized as follows: Section 2 reviews the related literature on shape control methods in large scale part assembly. Section 3 introduces the proposed reinforcement learning framework for sequential actuator placement optimization. Section 4 provides case studies to validate our methodology. Finally, this article is concluded in Section 5.

## 2. Literature Review

In the field of large-scale compliant part assembly, shape control remains a critical challenge due to the compliant nature of aircraft fuselages and ship hull panels. Existing research can be broadly categorized into two streams: actuator placement methods for aircraft fuselage assembly and fixture layout methods for ship hull panel assembly. Actuator placement strategies often focus on optimizing positions and forces to minimize assembly gaps and improve dimensional accuracy, while fixture layout methods concentrate on determining optimal positions to ensure stability and precision during assembly. Despite different tasks, these two problems share commonalities in addressing the complexities of large and compliant parts. However, a significant gap exists in the literature regarding a unified framework that can simultaneously optimize both the number of actuators and the forces they apply, particularly in the context of fuselage shape control. This article addresses this gap by introducing a novel reinforcement learning framework designed for sequential decision-making in actuator placement and force computation, offering a more efficient and adaptable solution for aircraft fuselage assembly. In following Sections 2.1 and 2.2, we introduced the relevant work in actuator placement for fuselage assembly and fixture layout optimization with application for ship hull.

## 2.1 Actuator Placement Methods for Aircraft Fuselage Assembly

In the field of actuator placement, existing methodologies often rely on uniform distribution strategies, where actuators are positioned at equal intervals between adjacent points. While this approach is straightforward, it frequently leads to suboptimal outcomes, such as excessive force application at certain locations, which can compromise the efficiency of shape adjustments. Moreover, this inefficiency translates into higher operational costs and increased complexity during assembly. To mitigate these limitations, recent advancements in actuator placement strategies have focused on leveraging innovative computational techniques. For instance, surrogate model-based control strategies (Yue et al., 2018) and sparse learning models (Du et al., 2022; Du et al., 2019) have emerged as promising solutions to optimize actuator placement, thereby enhancing both performance and cost-effectiveness. Yue et al. (2018) presented an automated shape control system for composite parts in aerospace manufacturing, addressing limitations of current manual metrology methods. The authors develop a surrogate model incorporating various uncertainties and embed it into a feed-forward control algorithm, demonstrating significant improvements in assembly time and dimensional quality. Based the proposed surrogate model, Du et al. (2019) presented a novel method for determining the optimal locations of actuators used in adjusting the shape of composite fuselages. The authors developed a sparse learning model and an efficient algorithm integrating the Alternating Direction Method of Multipliers (ADMM) to achieve better shape control performance compared to traditional fixed placements. Aiming at minimizing the maximum dimensional gap between composite fuselages during assembly, Du et al. (2022) introduced a novel sparse-learning model. The authors proposed a sparse learning methodology that considers the initial gap between a pair of fuselages and optimizes the adjustment to an intermediate shape, rather than adjusting each fuselage to a fixed design shape for maximum gap reduction, thereby significantly reducing the maximum gap and improving assembly quality.

In recent years, Reinforcement Learning (RL) has received significant attention for its ability to achieve outstanding performance in games. Meanwhile, some studies have attempted to apply RL to solve optimization problems in engineering (Hu et al., 2017; Li et al., 2024). In the past few years, there have also been relevant research efforts in the fuselage actuator placement. Lutz et al. (2024) presented a novel model-free reinforcement learning approach for adaptive shape control of composite fuselages during aircraft assembly. Their method utilizes a reinforcement learning agent to directly adjust fuselage sections in response to part variations, significantly reducing the root-mean-square gap between sections and outperforming benchmark methods in terms of final shape gap and maximum forces applied. These studies all demonstrate the unique strengths of Reinforcement Learning (RL) when it comes to optimizing engineering problems. Specifically, RL excels in dynamic decision-making, adapting to unpredictable and interactive environments through trial and error. Unlike traditional machine learning methods that rely on static datasets, RL agents learn from experiences, much like humans do in real life. This capability allows RL to tackle complex optimization tasks that require continuous adaptation and improvement. Moreover, RL's ability to balance exploration and exploitation enables it to find optimal solutions efficiently. By exploring new actions and exploiting known rewarding actions, RL can effectively navigate the trade-off between trying new strategies and sticking to proven ones, leading to better outcomes in engineering optimization problems. The RL-based fuselage shape control method (Lutz et al., 2024) has certain limitations. It directly optimizes an 18-dimensional force vector, which restricts its flexibility in addressing broader optimization challenges. Specifically, the framework lacks adaptability for optimizing the number of actuators.

## 2.2 *Fixture Layout Methods for Ship Hull Panel Assembly*

In the field of ship hull manufacturing, the assembly of large thin-walled parts poses significant challenges due to their tendency to deform under gravity, which can affect the final product's quality. Several studies have focused on optimizing fixture layouts to minimize dimensional deviations and improve assembly efficiency. Here is a summary of the key relevant studies:

Liu et al. (2020) developed a hybrid nonlinear variation model to predict and control the deformation of compliant metal plate assemblies in shipbuilding, considering factors like welding shrinkage and angular distortion. This model aims to improve the accuracy of predicting assembly deviations. Du et al. (2021) proposed an optimal design methodology for fixture layouts in ship assembly processes, integrating the direct stiffness method and simulated annealing algorithm. This approach focuses on minimizing dimensional gaps along the assembly interface to enhance the quality and efficiency of seam welding. Hong et al. (2024). introduced a butt clearance control-oriented fixture layout optimization method for large compliant ship parts. The study emphasizes the importance of controlling butt clearance by optimizing fixture layouts, using a modified finite element equation and a constrained multi-objective integer nonlinear programming model.

Some studies have explored deep learning and reinforcement learning for this problem, aiming to further enhance the optimization and prediction capabilities in fixture layout design.

For example, Jin et al. (2023)presented a transformer-based surrogate model with two-stage Latin hypercube sampling to predict deformations of compliant parts in ship sub-assembly processes. This method aims to improve prediction accuracy and efficiency by considering fixture positions and deviations. Wang et al. (2024) proposed SmartFixture, a physics-guided reinforcement learning framework for automatic fixture layout design. This approach uses deep reinforcement learning to interact with finite element analysis simulations, optimizing fixture layouts to minimize shape deformations in manufacturing systems.

Most studies on fixture layout optimization design and actuator placement are carried out under the situation that the number of fixtures is known. In previous studies, the optimization of the fixture quantity predominantly relied on a trial-and-error approach. Engineers would incrementally add or subtract fixtures at each stage of the process. Subsequently, they would search for the most suitable fixture arrangement separately in each step. Eventually, they determined the minimum number of fixtures required for the task (Liu et al., 2024b). Liu et al. (2024a) focused on optimizing the number of fixtures in large thin-walled parts assembly. They developed an improved particle swarm optimization algorithm to minimize the number of fixtures while ensuring part deformation and assembly gaps meet required tolerances.

In the field of aircraft fuselage assembly, optimizing actuator placement and force computation remains a complex challenge. Previous studies have made significant progress in applying reinforcement learning (RL) to optimize engineering problems. However, there is still a lack of a unified framework that can effectively optimize the number of actuators while also addressing the specific requirements of fuselage shape control. Unlike fixture layout optimization, which focuses solely on determining optimal positions, fuselage shape control requires simultaneous optimization of both actuator positions and the corresponding forces they apply. This dual requirement adds a layer of complexity that existing methods, primarily designed for position optimization, have yet to fully address.

To bridge this gap, this article introduces a novel RL framework specifically designed for sequential decision-making in actuator placement selection and optimal force computation. The framework is meticulously designed to incrementally select actuators based on dynamic system states, allowing for a more adaptive and efficient optimization process. This framework possesses the capacity to significantly reduce the number of actuators utilized. This has the potential to cut down on costs and complexity, making it a more efficient and practical approach in aircraft fuselage assembly scenarios.

## 3. Methodology

In this work, we address the problem of optimizing the shape control of composite fuselages under the assumption of elastic deformation. We present a comprehensive methodology for addressing the actuator placement optimization problem. First, in 3.2 (Submodular Function), we transform the actuator placement optimization problem into a submodular function formulation, which provides new insights for analysis and solutions. This transformation allows us to take advantage of the unique properties of submodular functions and further convert it into a linear programming problem for easier handling. Subsequently, in 3.3 (Markov Decision Process), we define the Markov decision process for actuator selection. We detail how the state variables are defined, how the agent makes decisions, and how the environment responds and provides rewards. Additionally, we explain the adoption of the greedy policy by the agent and the necessary variable transformations. Finally, in terms of the Dueling Q-Network Agent, we construct a neural network agent based on the Dueling Double Deep Q-Network (D3QN) algorithm with justifications on algorithm selection, the input to the Dueling Q-Network, and the overall structure of the Dueling Q-Network, aiming to enhance the learning efficiency and agent performance for actuator selection and placement tasks. The framework of our sequential actuator placement optimization for fuselage assembly via reinforcement learning (SAPO-RL) is shown in Figure 3.

### 3.1 *Surrogate Model*

When controlling the shape of a composite fuselage, during the adjustment process, only elastic deformation exists. Based on the assumption of elastic deformation, it is nature that the mechanical behavior of the fuselage deformation is linear to actuator forces according to the principles of mechanics (Yue et al., 2018). As a result, the deviations in the adjusted shape can be expressed as

$$\boldsymbol{\delta} = \boldsymbol{\psi} + \boldsymbol{U}\boldsymbol{F}$$

$$s.t. \boldsymbol{F}_l \leq \boldsymbol{F} \leq \boldsymbol{F}_L. \tag{1}$$

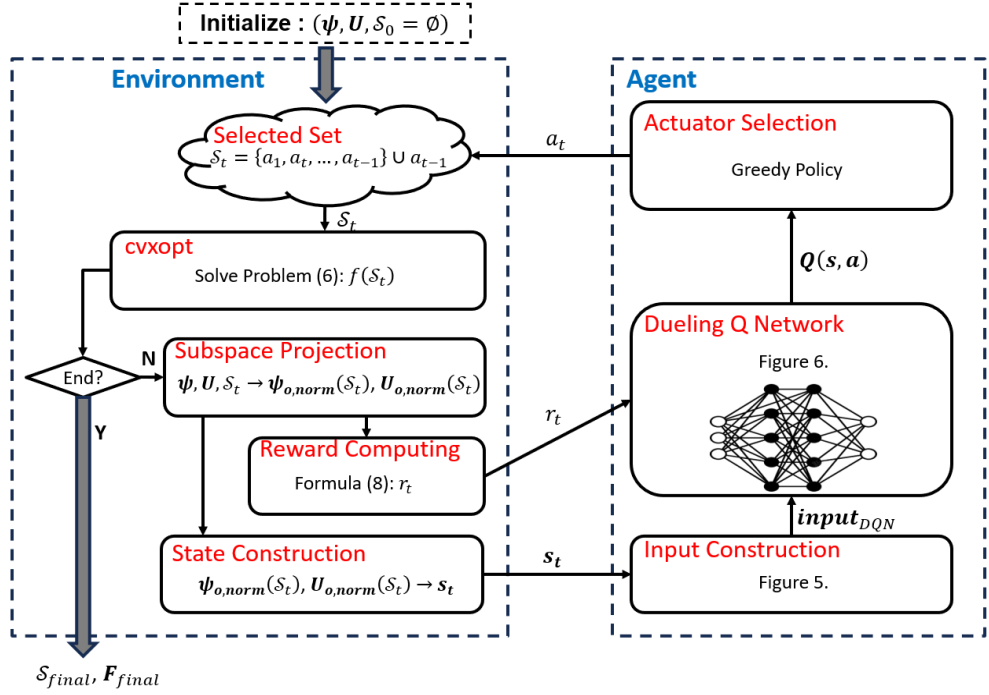

**Figure 3.** The framework of SAPO-RL

In the context of this model, $\boldsymbol{\delta} \in \mathbb{R}^n$ represents the final gap in the $Y$ and $Z$ directions following shape control. Specifically, this deviation refers to the disparity of the measurement point between two incoming fuselages at the same measurement position after the shape control, which is achieved through the application of actuator forces. Here, $n$ refers to the quantity of measurement points on the incoming fuselages. Meanwhile, $\boldsymbol{\psi} \in \mathbb{R}^n$ corresponds to the initial dimension deviations before the actuation forces are applied, and $\boldsymbol{U} \in \mathbb{R}^{n \times m}$ denotes the displacement matrix of the incoming fuselages. The variable $m$ stands for the number of available positions (for instance, the possible positions where an actuator can be installed) for the actuators. Additionally, $\boldsymbol{F} \in \mathbb{R}^m$ indicates the force exerted during the shape control operation. Furthermore, when carrying out the optimization, the safety requirements of the applied forces by actuators must be taken into account as well. $F_l$ and $F_L$ are the lower bound and the upper bound of actuators force vector $F$.

During the assembly of composite fuselages, the maximum gap, measured by the $L_\infty$ norm, is a critical factor affecting the assembly quality (Du et al., 2022). In this paper, we follow Du et al. (2022) and primarily focus on minimizing the maximum gap after adjustment. Consequently, our optimization objective function can be summarized as:

$$\min_F \|\boldsymbol{\delta}\|_\infty$$
$$s.t. \begin{matrix} \boldsymbol{\delta} = \boldsymbol{\psi} + \boldsymbol{UF} \\ F_l \leq F \leq F_L \end{matrix}. \qquad (2)$$

In practical applications, the number of available actuators is often restricted. Therefore, out of numerous possible actuator positions, we aim to identify the most efficient ones. The optimal actuator positions are those associated with the non-zero elements of the force vector. There are $m$ viable actuator positions along a pair of fuselages. However, in actual shape control procedures, only a total of $M$ actuators are utilized. Consequently, the number of non-zero elements in the force vector are the exact number of $M$, which can be formulated as the following problem:

$$\min_F \|\boldsymbol{\delta}\|_\infty$$
$$s.t. \begin{matrix} \boldsymbol{\delta} = \boldsymbol{\psi} + \boldsymbol{UF} \\ F_l \leq F \leq F_L \\ \|F\|_0 = M \end{matrix}. \qquad (3)$$

## 3.2 *Submodular Function*

It can be noted that Problem (3) can actually be transformed into a submodular function form (Nemhauser et al., 1978) as follows:

$$\min_{\mathcal{S}} f(\mathcal{S})$$
$$s.t. \begin{matrix} |\mathcal{S}| \leq M \\ \mathcal{S} \subseteq \mathcal{E} \end{matrix}, \qquad (4)$$

where $\mathcal{S}$ represents the set of the indicators for selected actuators, $|\mathcal{S}|$ denotes the cardinality of the set $\mathcal{S}$, which represents the number of elements in the set, and $\mathcal{E}$ is the set of positions of all actuators. $f(S)$ is defined as follows:

$$f(S) = \min_{F_S} \|\delta\|_\infty$$

$$\delta = \psi + U_S F_S$$
$$s.t. \ F_l \leq F_S \leq F_L \ . \tag{5}$$
$$|S| \leq M$$

In problem (5), $U_S \in \mathbb{R}^{n \times |S|}$ is the displacement matrix composed of the displacement vectors of the selected actuators, and $F_S \in \mathbb{R}^{|S|}$ is the force applied by the selected actuators. This transformation provides a new perspective for analyzing and solving the problem, enabling us to take advantage of the unique properties of submodular functions. Submodular functions have been widely studied in various fields due to their desirable characteristics such as the diminishing returns property. By reformulating Problem (3) in this way, we can potentially apply well-established theories related to submodular functions. For Problem (5), it is quite straightforward to transform it into a linear programming problem for solution. Linear programming offers a well-established framework with a variety of efficient algorithms, which can significantly simplify the problem-solving process. The transformed linear programming problem is presented as follows:

$$f(S) = \min_{F_S} d$$

$$s.t. \ \begin{matrix} U_S F_S + d\mathbf{1} + \psi \geq 0 \\ U_S F_S - d\mathbf{1} + \psi \leq 0 \\ F_S - F_L \leq 0 \\ F_S - F_l \geq 0 \end{matrix} \ . \tag{6}$$

### 3.3 *Markov Decision Process*

The Markov decision process for actuator selection is as designed follows. In the initial state, the set of selected actuators $S_0$ is an empty set. The state variables $s_t$ consist of the initial deviation $\psi$, the displacement matrix $U$, and the set of selected actuators $S_t$. The Agent calculates and outputs the decision based on the state, where it selects the position of the next actuator to apply force from the set of unselected actuators $a_t$ and passes to the environment. Upon receiving, the environment first updates the set of selected actuators. Then, according to the current selected actuators $\mathcal{S}_t$, The linear programming solver (use cvxopt package in our python implementation) solves Problem (6) and calculates the minimum adjusted maximum gap achievable by the currently selected set of actuators. Finally, the environment calculates the reward based on the result, updates $s_{t+1}$ and $t$, and passes them to the Agent for the next decision until $|\mathcal{S}| = M$. The described Markov Decision Process is illustrated in Figure 4.

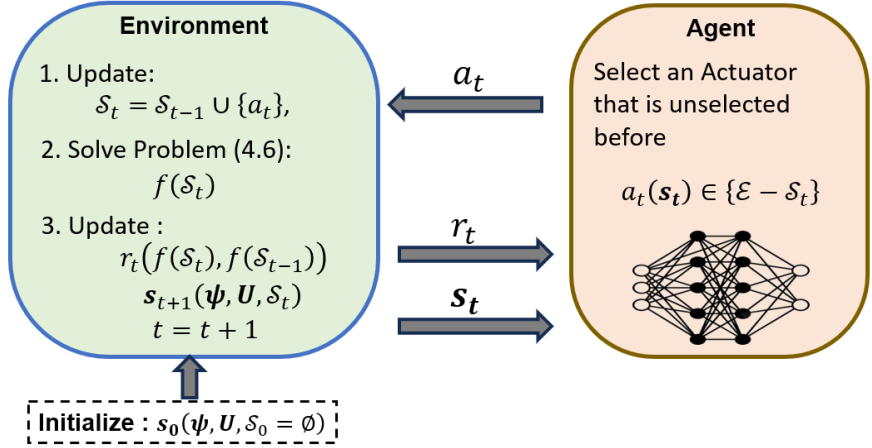

**Figure 4.**  Markov decision process for sequential actuator placement selection.

Since we have modeled the problem as a submodular optimization problem, greedy policy is a commonly used method to solve such optimization problem. Greedy policy has the advantage as it is simple and easy to implement. In addition, such method can achieve results close to the optimal solution in many cases such as the set-covering problem (Nemhauser et al., 1978). Therefore, our agent also adopts the greedy policy, that is, in each selection, the position of the actuator will be selected with the highest reward. Given this policy, agent action $a_t$ can be solved from the following formula:

$$a_t = \underset{e \in \{\mathcal{E}-\mathcal{S}_t\}}{\mathrm{argmax}}[f_{U,\psi}(\mathcal{S}_t) - f_{U,\psi}(\mathcal{S}_t \cup \{e\})] \tag{7}$$

In Problem (7), the input variables include sets, which makes it difficult to solve. Due to the lack of order and fixed structure of sets, non-differentiable set operations, and the lack of a natural metric for similarity, it becomes difficult to calculate the gradients when used as input variables for functions or neural networks. Therefore, we make the following transformation to Problem (7):

$$\underset{e\in\{\mathcal{E}-\mathcal{S}_t\}}{\operatorname{argmax}}[f_{U,\psi}(\mathcal{S}_t) - f_{U,\psi}(\mathcal{S}_t \cup \{e\})] = \underset{e\in\{\mathcal{E}-\mathcal{S}_t\}}{\operatorname{argmax}}[f_{U_o(\mathcal{S}_t),\psi_o(\mathcal{S}_t)}(\{e\})]$$

$$= \underset{e\in\{\mathcal{E}-\mathcal{S}_t\}}{\operatorname{argmax}} g_{\psi_o(\mathcal{S}_t)}[\boldsymbol{u}_o^e(\mathcal{S}_t)], \tag{8}$$

where $f_{U,\psi}(\cdot)$ denote the solution of Problem (6) when the displacement matrix is $U$ and the initial deviation is $\psi$. $U_o(\mathcal{S}_t)$ represents a new displacement matrix composed of the orthogonal vectors obtained by projecting the vectors in $U$ onto the subspace $U_{\mathcal{S}_t}$. $\psi_o(\mathcal{S}_t)$ is the orthogonal vector obtained by projecting $\psi$ onto the subspace $U_{\mathcal{S}_t}$. $\boldsymbol{u}_o^e(\mathcal{S}_t)$ stands for the $e$-th row vector in $U_o(\mathcal{S}_t)$. It is obvious that if $e \in \mathcal{S}_t$, that is, if an actuator that has already been selected is chosen again, then $\boldsymbol{u}_o^e(\mathcal{S}_t)$ is an zero vector and $g_{\psi_o(\mathcal{S}_t)}[\boldsymbol{u}_o^e(\mathcal{S}_t)] = f_{U,\psi}(\mathcal{S}_t) - f_{U,\psi}(\mathcal{S}_t) = 0$.

There is a frequent occurrence when addressing submodular problems with reinforcement learning methods that rewards naturally have diminishing returns. In our research, not only does the reward exhibit diminishing returns, but the norms of $\psi_o(\mathcal{S}_t)$ and $U_o(\mathcal{S}_t)$ also significantly decrease as $t$ increases (Prajapat et al., 2023). To prevent this problem, which could lead to difficulties in gradient calculation, we perform $L_2$-normalization on $\psi_o(\mathcal{S}_t)$ and non-zero row vectors of $U_o(\mathcal{S}_t)$ to get $\psi_{o,norm}(\mathcal{S}_t)$ and $U_{o,norm}(\mathcal{S}_t)$. Simultaneously, we design the reward for each step as follows:

$$r_t = \frac{f_{U,\psi}(\mathcal{S}_t) - f_{U,\psi}(\mathcal{S}_t \cup \{a_t \in \{\mathcal{E}-\mathcal{S}_t\}\})}{\|\boldsymbol{\delta}_{\mathcal{S}_t}\|_2}, \tag{9}$$

where $\boldsymbol{\delta}_{\mathcal{S}_t}$ is the gap vector obtained from the solution $f_{U,\psi}(\mathcal{S}_t)$ of Problem (6). $f(\mathcal{S}_t) - f(\mathcal{S}_t \cup \{a_t \in \{\mathcal{E}-\mathcal{S}_t\}\})$ represents the reduction in maximum gap after adding the actuator indicated by $a_t$.

Therefore, we design the structure of the state $\boldsymbol{s}_t$ as a $(m+1) \times (n+1)$ matrix. In this matrix, the first $m$ rows and $n$ columns represent $U_{o,norm}(\mathcal{S}_t)$. The first $n$ elements of the $(m+1)^{th}$ row are $\psi_{o,norm}(\mathcal{S}_t)$. The $(n+1)^{th}$ column records the selection status of the actuators, which serves as a mask when the agent selects an actuator. This design allows the agent to better utilize the information related to the displacement matrix, initial deviation, and the selection history of actuators during the decision-making process. This design of the state $s_t$ provides a more organized and efficient way for the agent to process information and make optimal choices based on the current state.

### 3.4 *Dueling Q-Network Agent*

In this work, we construct a neural network agent. As previously mentioned, our agent will also adopt the greedy policy. To enhance the learning efficiency and performance of the agent, we choose to implement the D3QN algorithm due to its combination of Dueling DQN and Double DQN. This combination leverages the Dueling DQN's ability to independently estimate state and action values, improving the accuracy of Q-value estimation, while the Double DQN component addresses the overestimation problem of Q-values by separating action selection and evaluation. D3QN also maintains the greedy policy's efficiency in action selection while mitigating the issue of overestimation common in other Q-value-based algorithms.

The input to the Dueling Q-Network is a matrix $input_{DQN} \in \mathbb{R}^{m \times 2n}$, which is reconstructed from $s_t$. As shown in Figure 5, $\psi_o(S_t)$ is replicated $m$ times and concatenated behind $U_o(S_t)$. This enables the Agent to effectively utilize the information in the state $s_t$ and better learn $g_{\psi_o(S_t)}[u_o^e(S_t)]$, improving the accuracy and efficiency of the agent's decision-making process in the context of the proposed problem setup. The structure of the Dueling Q-Network is shown in Figure 6.

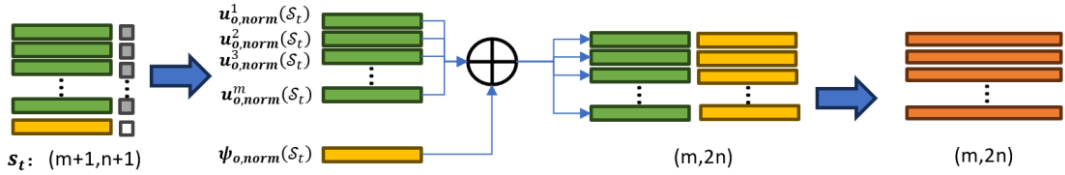

**Figure 5.**  Dueling Q-Network input variable construction.

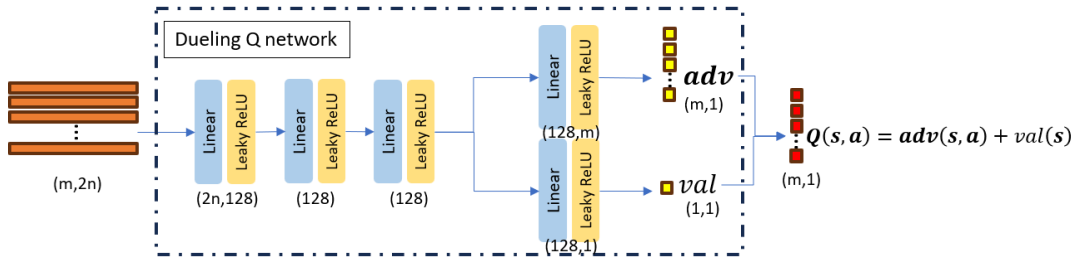

**Figure 6.**  Dueling Q-Network for sequential actuator placement selection.

# 4. Case Study

In this section, we conduct a series of experiments to evaluate the performance and capabilities of our proposed methodology. Specifically, Subsection 4.2 focuses on comparing our SAPO-RL method with the PPO method from Lutz et al. (2024) to demonstrate the effectiveness of our approach in reducing assembly gaps. Subsection 4.3 presents an ablation study between SAPO-RL and its variant SAPO-ReEs to evaluate the impact of the reinforcement learning component. Finally, Subsection 4.4 explores the optimization of the actuator number using our framework, highlighting its flexibility and adaptability in meeting different manufacturing requirements.

## 4.1 *Data Description*

Our experiment utilized the publicly available data and framework presented in the paper by Lutz et al. (2024). In the experiment, the dimension of the deviation $n = 354$, the number of positions where actuators can be placed $m = 18$, and the number of positions to be selected for placing actuators is $M = 10$. The objective of optimization is to minimize the maximum gap (MG). We trained our agent using 50 pairs of fuselages and tested with 50 pairs of fuselages and compared it with the agent in Lutz et al. (2024), which was also trained and tested on the same pairs of fuselages. Due to significant differences in the environment and reward designs between the methods, there was a large disparity in the single-step computation time. To ensure a fair comparison, we controlled the total training time for all three methods to be approximately 40–50 minutes. Specifically, SAPO-RL and SAPO-ReEs were trained for 100,000 steps (10000 episodes), while PPO, due to its inherently faster computation per step, was trained for 10,000,000 steps (10,000,000 episodes). This allowed us to evaluate the performance of each method under comparable training durations, despite differences in their computational complexities. The comparison metrics were the average maximum gap (MG) and root mean square gap (RMSG) values returned at the end of each episode of the training pairs and testing pairs during the training process.

## 4.2 Comparison with PPO (from Lutz et al. (2024))

This subsection compares our SAPO-RL method with the PPO method introduced by Lutz et al. (2024). During the training phase, the main goal is to explore the environment and learn the optimal policy. SAPO-RL employed an $\varepsilon$-Greedy policy to balance exploration and exploitation. With $\varepsilon = 0.1$, the agent has a 10% chance to select a random action (exploration) and a 90% chance to choose the action with the highest estimated value (exploitation). This helps to prevent the agent from getting stuck in local optima by occasionally trying out new actions, which is crucial for effective learning. Figure 7 shows the changes in MG and RMSG of 50 training pairs of fuselages over the training episode during the training process of the two methods.

For testing fuselages (testing phase), the primary objective is to evaluate the agent's performance using the knowledge it has acquired during training. At this point, further exploration is unnecessary, as the agent should already have learned the optimal actions. Using a pure Greedy policy ensures that the agent always selects the action it believes will yield the highest reward based on its learned policy. This provides a clear and stable assessment of the agent's performance, as it is not distracted by random actions. Figure 8 shows the changes in the average MG and RMSG of 30 testing pairs of fuselages over the training episode during the training process of the two approaches.

During the training process, it can be observed that both MG and RMSG of the two methods tend to decrease over time, demonstrating that all methods enable the agent to learn how to apply appropriate forces. Notably, since our SAPO-RL method incorporates a linear programming solution module into the environment, these agents were able to apply relatively more suitable forces compared to PPO from the beginning of the training.

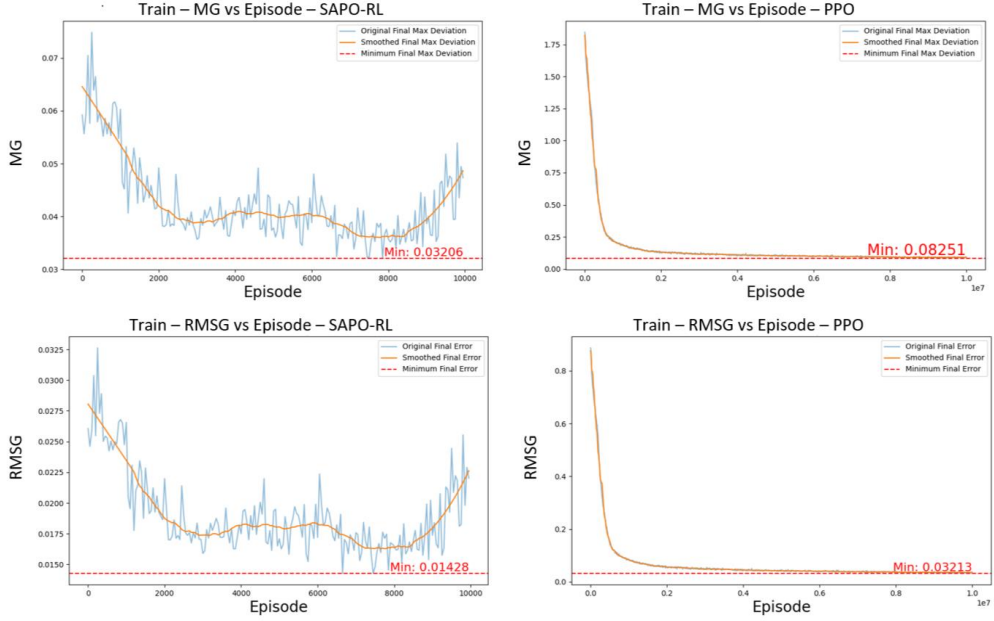

**Figure 7.**  MG and RMSG of 50 training pairs (SAPO-RL and PPO).

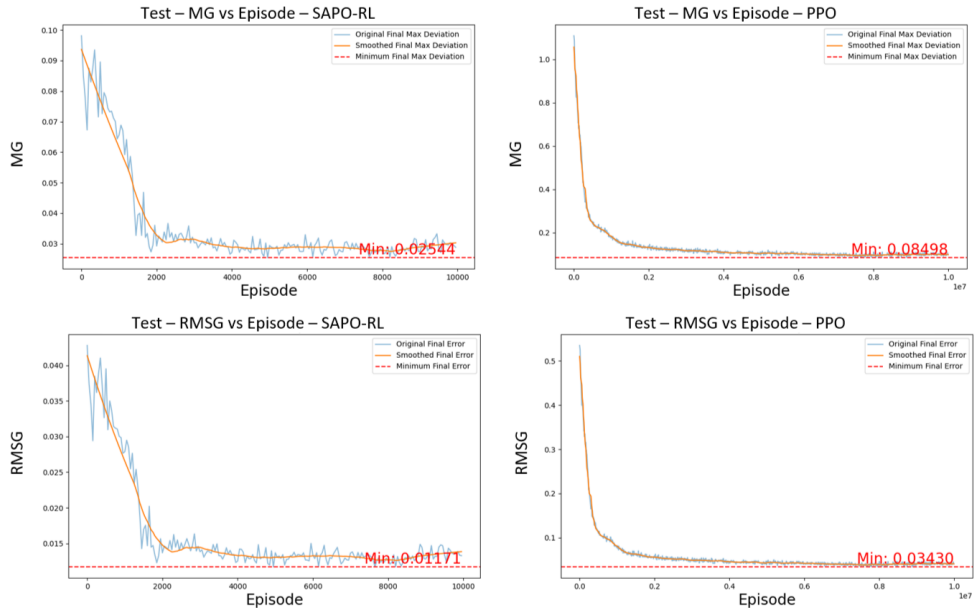

**Figure 8.**  MG and RMSG of 30 testing pairs (SAPO-RL and PPO).

In the later stage of training, MG and RMSG achieved by our SAPO-RL method were smaller than those of the PPO method in Lutz et al. (2024). This is partly due to the training setup where SAPO-RL was trained for 100,000 steps (10000 episodes), while PPO, benefiting from faster computation per step, was trained for 10,000,000 steps (10,000,000 episodes). Despite the difference in steps, the training durations were comparable, and our methods maintained computational efficiency. Notably, SAPO-RL incorporated a linear programming solution module into the environment from the start, allowing them to apply more suitable forces compared to PPO. The comparison metrics, average MG and RMSG, consistently showed SAPO-RL's effectiveness in improving assembly quality while keeping computational efficiency.

We applied the final output forces provided by the agents for 30 testing pairs of fuselages to corresponding fuselage models in the ANSYS Simulator. The finite element analysis (FEA) results, including the average of MG and RMSG for these 30 testing pairs of fuselages, are listed in Table 1. It can be observed that SAPO-RL outperforms PPO. Examples of actuator selections and corresponding forces for the two approaches (SAPO-RL and PPO) are illustrated in Figure 9. The figure shows the cross-sectional view of the fuselage joining edge before and after shape control has been applied.

**Table 1.** MG and RMSG of 30 testing pairs in final ANSYS validation.

| Index | SAPO-RL | PPO |
|---|---|---|
| **MG (Mean)** | **0.0323** | 0.0500 |
| **RMSG (Mean)** | **0.0141** | 0.0156 |

## 4.3 *Ablation Study (SAPO-RL vs. SAPO-ReEs)*

This subsection presents an ablation study comparing SAPO-RL with its variant, SAPO-ReEs, to evaluate the effectiveness of the reinforcement learning (RL) component in SAPO-RL. Both methods were trained under the same experimental setup as introduced in Subsection 4.2, with the objective of minimizing the maximum gap (MG) and root mean square gap (RMSG).

The SAPO-ReEs method combines a deep neural network (DNN) for reward estimation with a greedy policy for actuator placement selection. This approach leverages the DNN to predict the reward associated with each possible actuator position at each step and then applies a greedy policy to select the actuator position that maximizes the predicted reward. The input to the Reward Estimation network is a vector of length $(2n)$, which is reconstructed from $s_t$, as shown in Figure 5. The structure of the Reward Estimation network is shown in Figure 10.

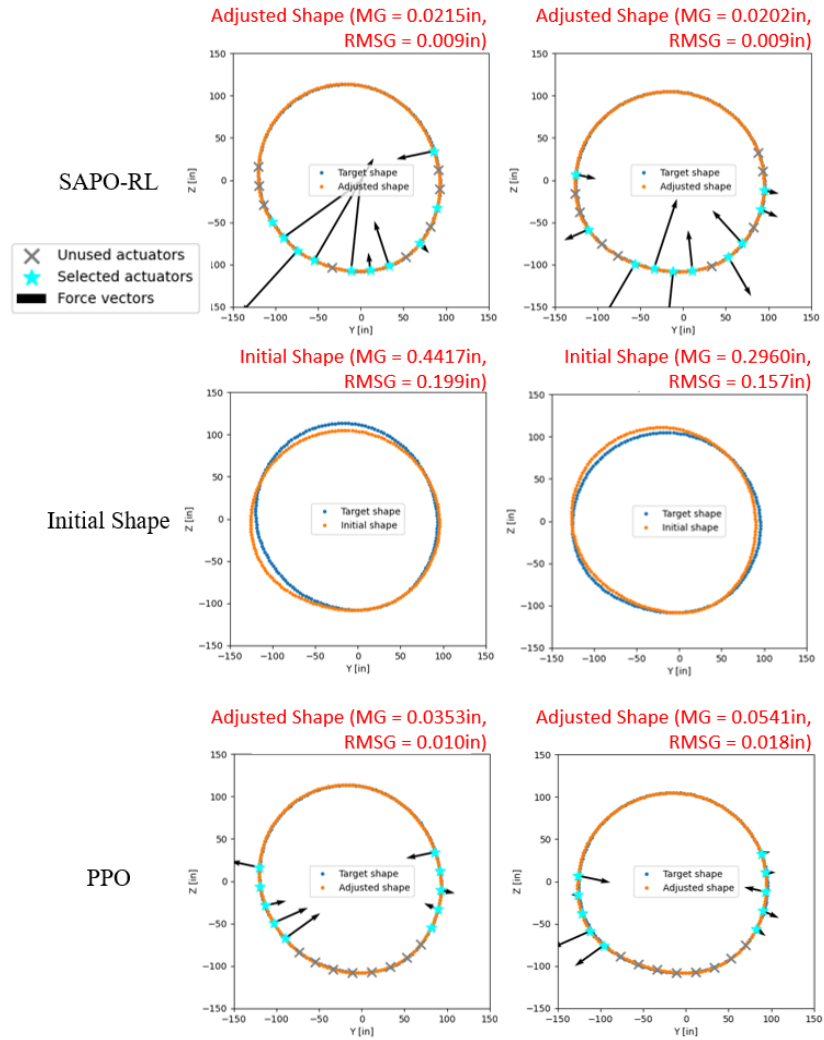

**Figure 9.** Examples of shape adjustment results via ANSYS validation (SAPO-RL and PPO).

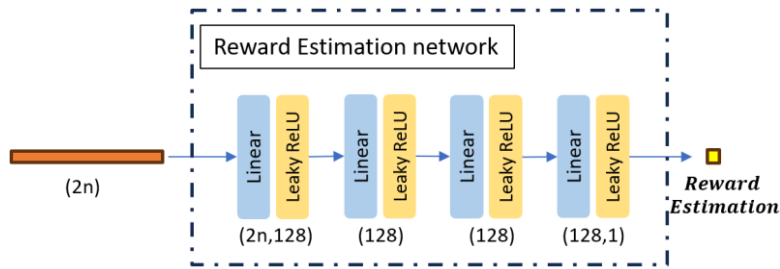

**Figure 10.** Reward Estimation network for sequential actuator placement selection.

Figure 11 illustrates the changes in the average MG and RMSG of 50 training pairs of fuselages over the training episode. Figure 12 shows the changes in the average MG and RMSG of 30 testing pairs of fuselages over the training episode during the training process of the two approaches. The results show that SAPO-RL achieved better performance in reducing both MG and RMSG compared to SAPO-ReEs. This highlights the effectiveness of using the SAPO-RL reinforcement learning method, which can estimate both current and future rewards through its advanced Q-learning framework, rather than relying solely on a greedy policy.

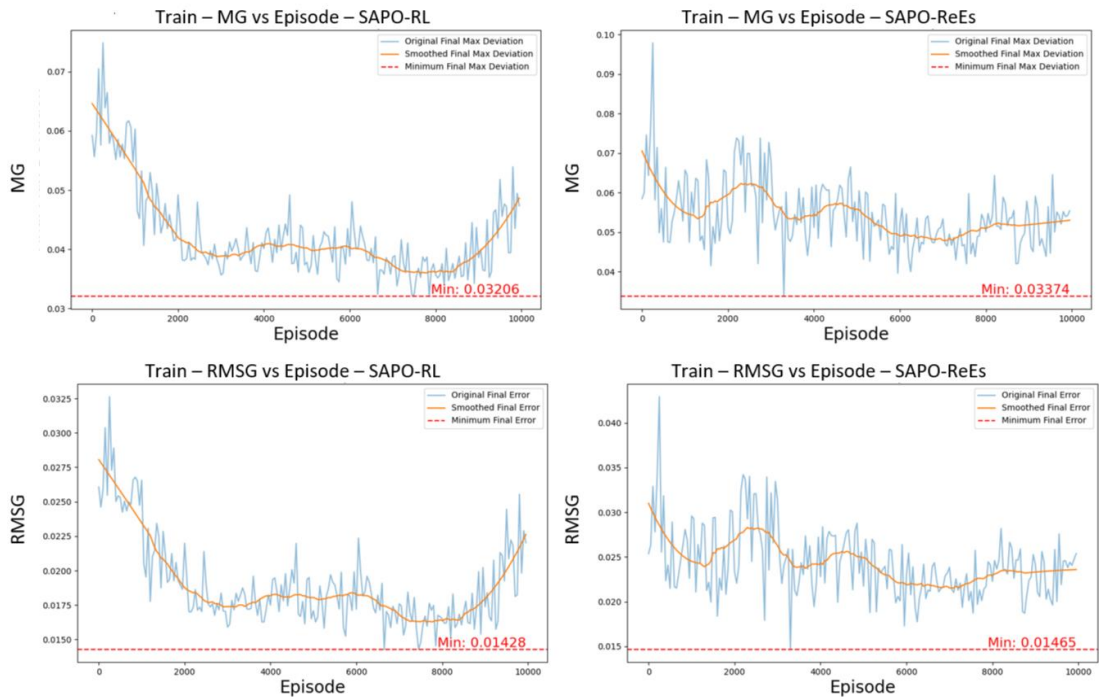

**Figure 11.** MG and RMSG of 50 training pairs (SAPO-RL and SAPO-ReEs).

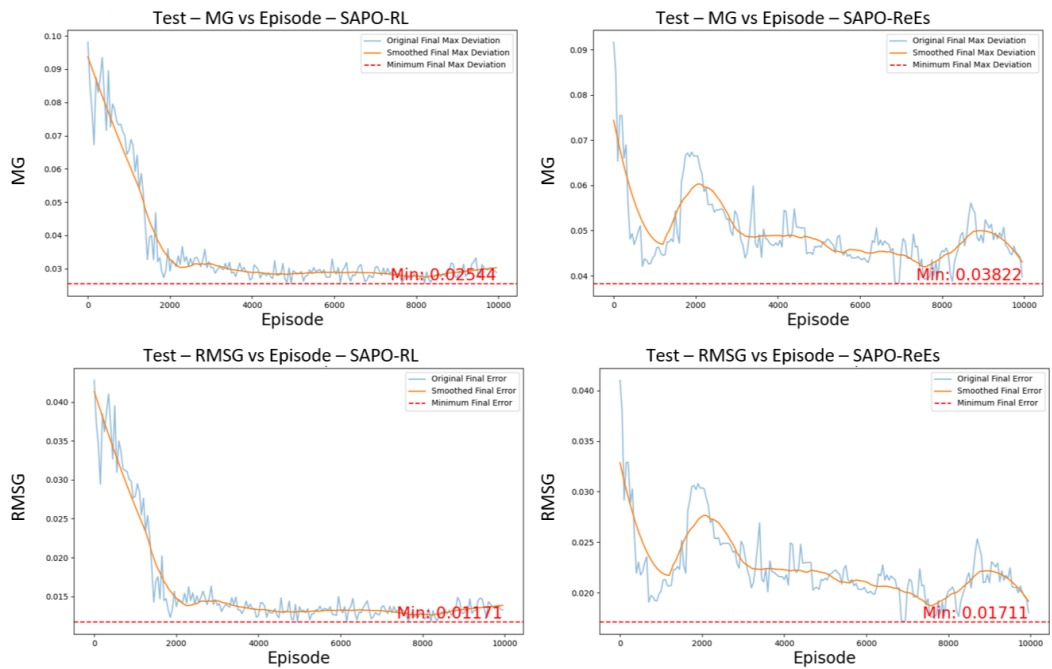

**Figure 12.** MG and RMSG of 30 testing pairs (SAPO-RL and SAPO-ReEs).

We applied the final output forces provided by the agents for 30 testing pairs of fuselages to corresponding fuselage models in the ANSYS Simulator. The finite element analysis (FEA) results, including the average of MG and RMSG for these 30 testing pairs of fuselages, are listed in Table 2.

**Table 2.** MG and RMSG of 30 testing pairs in final ANSYS validation.

| Index | SAPO-RL | SAPO-ReEs |
|---|---|---|
| **MG (Mean)** | **0.0323** | 0.0420 |
| **RMSG (Mean)** | **0.0141** | 0.0181 |

Examples of actuator selections and corresponding forces for the two approaches (SAPO-RL and SAPO-ReEs) are illustrated in Figure 13. The figure shows the cross-sectional view of the fuselage joining edge before and after shape control has been applied. The actuator selections via the two methods have a high degree of overlap, indicating that both agents have learned similar policies for actuator placement. However, the policy learned by SAPO-RL is slightly more effective, as evidenced by its superior performance in reducing the MG and RMSG. This further underscores the advantages of SAPO-RL in achieving higher precision in fuselage assembly.

### 4.4 *Optimization Results of Actuator Number*

Compared to Lutz et al. (2024), our reinforcement learning framework, which sequentially selects actuators, can optimize the number of actuators by simply modifying the episode termination criteria. This demonstrates the applicability and flexibility of our framework.

After the training, for the 30 testing pairs of fuselages, we modified the episode ending criterion of the environment to end when $MG < Limit_{MG}$ and recorded the final number of actuators. We conducted experiments with six different MG Specifications, i.e., 0.025, 0.03, 0.035, 0.04, 0.045, 0.05.

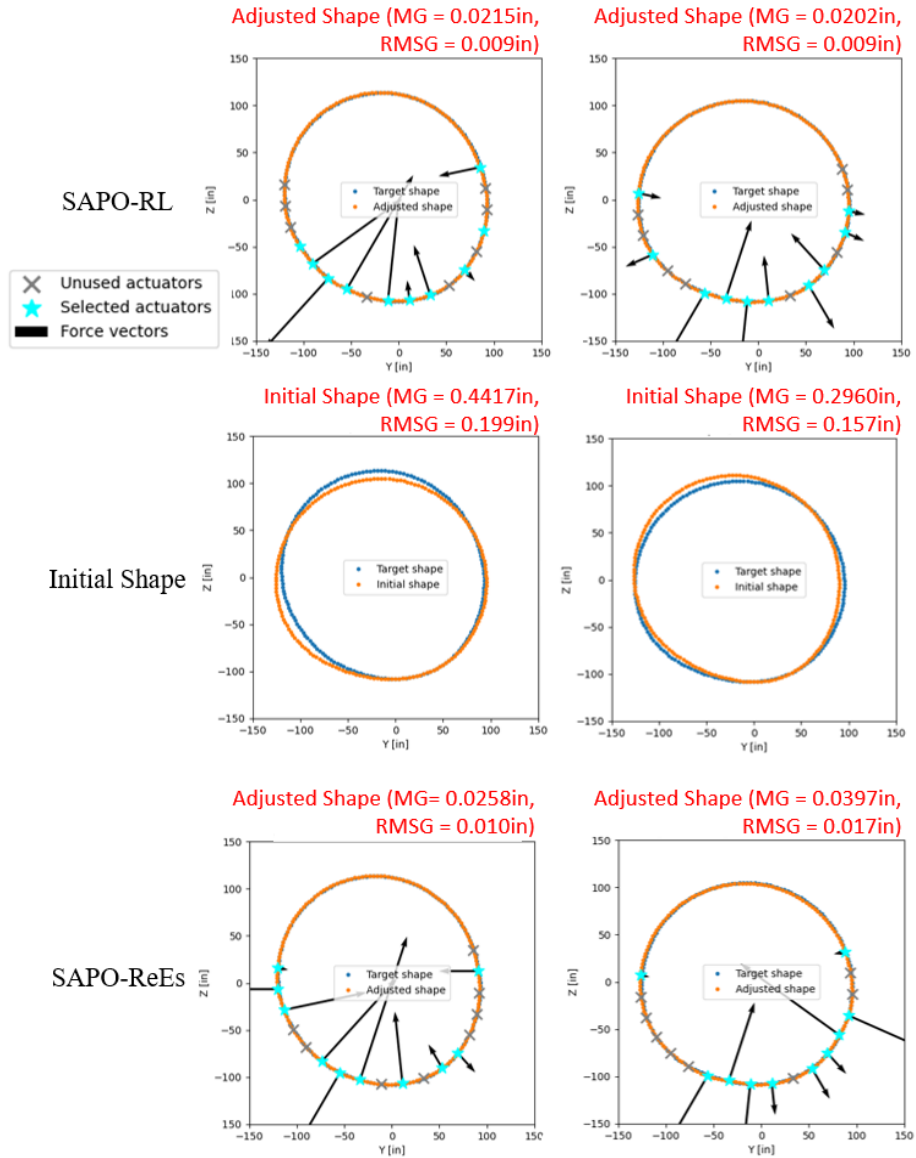

**Figure 13.** Examples of shape adjustment results via ANSYS validation (SAPO-RL and SAPO-ReEs).

The final selected number of actuators for these 30 testing pairs of fuselages under the six different specifications is shown in the Figure 14. As the specification value increases, the selected number of actuators decreases, achieving the goal of optimizing the number of actuators while meeting the manufacturing requirements, which reflects the good adaptability of our framework.

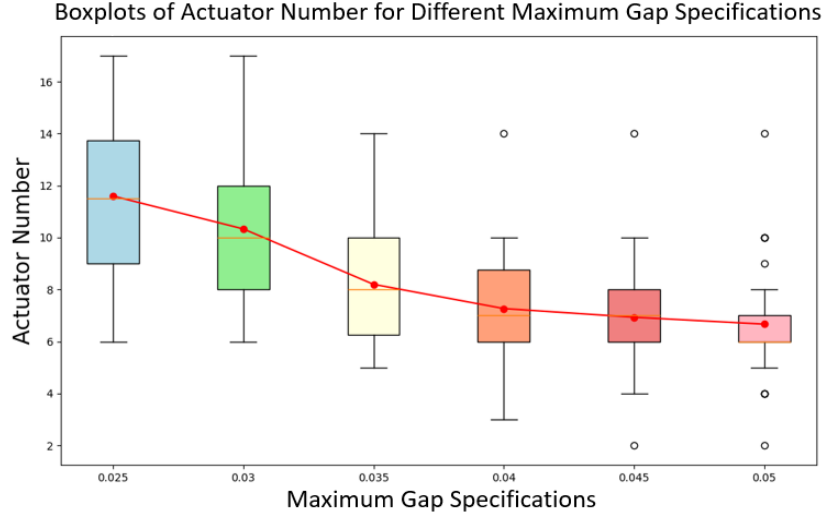

**Figure 14.**    Boxplots of the actuator number for the 30 testing pairs of fuselages.

## 5 Conclusion

The precise assembly of composite fuselages is of utmost significance for aircraft fuselage assembly. This study introduced a reinforcement learning (RL) framework SAPO-RL, centered around the Dueling Double Deep Q-Network (D3QN) algorithm, to tackle the challenges in actuator placement selection and optimal force allocation for composite fuselage assembly. The proposed methodology reformulates the actuator selection problem as a submodular function optimization problem, which not only facilitates the utilization of sub-modularity properties to obtain near-optimal solutions efficiently but also enhance computational efficiency. The Markov decision process was meticulously designed for sequential actuator selection. The agent, adopting a greedy policy, makes decisions based on the state variables, which include initial deviation, displacement matrix, and the set of selected actuators. The neural network agent, implemented with the D3QN algorithm, further improved learning efficiency and decision-making accuracy.

Notably, the sequential nature of our approach offers inherent flexibility to adapt to diverse optimization objectives without modifying the framework. For instance, the termination criteria for an episode—traditionally defined by reaching a fixed number of actuators—can be redefined to halt when the Max Gap (MG) falls within a specified threshold. This adjustment allows the system to dynamically optimize the number of actuators while ensuring compliance with manufacturing precision requirements. Such a modification not only reduces actuator usage and associated costs but also maintains the integrity of the original methodology, demonstrating its scalability and adaptability to varying operational constraints. Through numerical case studies and comparison studies, the effectiveness of the proposed method was demonstrated.

While our proposed methodology has achieved significant results in composite fuselage assembly optimization, there remain several promising avenues for further enhancement. The current framework, although robust and flexible, could benefit from improvements that address more complex assembly scenarios and tighter precision requirements. Additionally, as advanced manufacturing technologies continue to evolve, integrating our approach with these innovations holds the potential to unlock even higher levels of assembly precision and efficiency. Future work will focus on further improving the algorithm's performance, exploring its application in more complex assembly scenarios, and integrating it with other advanced manufacturing technologies to achieve even higher-precision assembly.

## Code Availability

The code will be released upon publication.

## Acknowledgments